# Critical bending and magnetic shape memory effect in magnetoactive elastomers

V. M. Kalita[1,2,3,4], Yu. I. Dzhezherya[2,3], S. V. Cherepov[2], Yu. B. Skirta[2], A. V. Bodnaruk[4], G.G. Levchenko [1,5]

[1] State Key Laboratory of Superhard Materials, International Centre of Future Science, Jilin University, Changchun 130012, China
[2] Institute of Magnetism of the NAS of Ukraine and MES of Ukraine, 36-b Vernadsky Blvd., Kyiv 03142, Ukraine
[3] National Technical University of Ukraine "Igor Sikorsky Kyiv Polytechnic Institute", ProspektPeremohy 37, Kyiv 03056, Ukraine
[4] Institute of Physics, NAS of Ukraine, ProspektNauky 46, Kyiv 03028, Ukraine
[5] Donetsk Institute for Physics and Engineering named after O.O. Galkin, NAS of Ukraine, ProspektNauky 46, Kyiv 03028, Ukraine

E-mail: g-levch@ukr.net

**Abstract**

The results of a study of magnetoactive elastomers (MAEs) consisting of an elastomer matrix with embedded ferromagnetic particles are presented. A continuous critical bending induced by the magnetic field, characterized by a critical exponent for the bending magnitude, and the derivative of which has a singularity in the critical region is reported for the first time. The mechanical stability loss and the symmetry reduction of the magnetic state, which are interrelated with each other, take place at the critical point. The magnetization in the high-symmetric state (below the critical point) is directed along the magnetic field and the torque is absent. Above the critical point, the magnetization and the magnetic field are noncollinear and there arises a torque, which is self-consistent with the bending. The magnetic field dependence of the MAE bending was found to have a hysteresis, which is associated with the magneto-rheological effect. The shape memory effect was also obtained for the MAE bending in a cycle consisting of magnetization, cooling (at $H\neq 0$), and heating (at $H=0$). The influence of the critical glass transition temperature of the matrix, as well as its melting/solidification temperature, on the magnetic shape memory effect was studied.

## 1. Introduction

Magnetoactive elastomers (MAEs) or magneto-rheological elastomers are composites consisting of an elastomer matrix with embedded magnetic filler particles. The MAEs exhibit large magneto-rheological [1-7] and magneto-dielectric [8-11] effects. They possess large magnetostriction (up to several tens of percent) [12-17] and undergo a huge bending in low magnetic fields [18-24]. The MAEs also demonstrate a shape memory effect [13,15,21,25-28] when a sample is elongated in a stationary magnetic field or heated in a small-amplitude ac magnetic field. In [28], the MAE shape memory effect was explained by strong pseudo-plasticity of the MAE in the magnetic field.

The unique properties of MAEs are of interest when the latter are applied as active dampers allowing a contactless control by means of a magnetic field [29-32], as stress or strain transducers also controllable by a magnetic field [33], as well as sensors or actuators [34-38]. For robotics [39-43], the shape memory phenomenon in MAEs and the deformations induced in them by magnetic field are of interest.

The anomalous properties of MAEs are associated with a possibility of filler restructuring in those media [44-47], namely, with the ability of filler particles to change their position and orientation in the course of MAE magnetization [48-52]. Due to the restructuring induced by a magnetic field, the elastic properties of MAEs strongly depend on this field. This is the so-called magneto-rheological effect [44-52]. The magnetic-field-induced reversible restructuring is most pronounced in the MAEs with elastically soft matrices [53]. In [28], pseudo-plasticity was also explained as a result of the MAE restructuring, so that it is a consequence of relaxation occurring at the MAE deformation. In this work, we will show that the restructuring is the main origin of the MAE bending hysteresis in the magnetic field.

The magnetic-field-induced critical bendingof the MAE beam was first reported only two years ago in Ref. [54], where a jump-like bending was discussed, and the critical field was considered as the field of stabilityloss.Earlier [55], the threshold and the loss of stability were observed when the MAE membrane was being magnetized. In [54], as well as in [55], when determining the critical field, a competition between the elastic and magnetostatic energies was taken into account. In [55], the membrane bending was explained by the magnetostriction effect associated with a change in the sample shape [56,57].Here, we report for the first time about a continuous critical bending induced by a magnetic field, for which the derivative of the function characterizing the bending rather than the function itself has a singularity, and this singularity is associated with a change in the magnetic state symmetry.

Modern directions of MAE applications(astronautics, medicine, robotics) make the investigations of the shape memory effect in MAEschallenging [13,15,21,25-28]. Therefore, much attention is paid to the detection and study of this effect in the elastomer under study. We show that the magnetic shape memory of MAEs is associated with the processes of glass transition and melting/solidification [58-60] in the polymer matrix, but not with the intrinsic magnetostriction of ferromagnetic filler particles in the MAE.The thermo-magnetic mechanism of the magnetic shape memory in MAEs was considered in [42]. In particular, the sample was first heated up using an ac field to a temperature above the glass transition temperature of the matrix and then the heated sample was subjected to bending by applying an external magnetic field. Afterward, the sample was cooled down and the external stationary magnetic field was turned off. The sample deformation created under the action of the field survived at that. We will show that not only the critical glass transition temperature but also the critical melting/solidification temperature of the matrix are important factors for the magnetic shape memory effect to reveal itself [58-60].

If an MAE possesses the magnetic shape memory [42], it can be heated up at the expense of magnetic losses by making use of an ac magnetic field with a small amplitude [27,61,62]. Therefore, we will study MAEs with weakly coercive particles and low residual magnetization.This circumstance may be of practical interest when creating devices in which the magnetic shape memory effect in MAEs is applied, especially for medical applications [63], when only contactless heating can be used.

In Ref. [64], it was experimentally shown that if the MAE is heated up making use of a small-amplitude ac magnetic field, the shape memory effect can take place provided that the field amplitude exceeds a certain threshold value, the origin of which has not been elucidated yet. Recently [65], it has been found that on the basis of the partial hysteresis loops, the caloric effect can be estimated. Therefore, to explain the MAE heating threshold, we measured the partial loops of the examined sample.

For practical application, it is important to possess the MAEs with different coexisting useful properties.

Thus,in this paper, following [42], we performed the comprehensive study of the magneto-mechanical features of the MAEs and their magneto-hysteretic properties.We showed that (i) the bending of an MAE critically depends on the magnitude of the uniform stationary magnetic field, (ii) the bending of an MAE in a uniform stationary magnetic field has the magnetic shape memory effect depending on both the glass transition and melting/solidification temperatures of the MAE matrix, and (iii) the heating-up threshold for the MAE in an ac magnetic field depends on the coercivity of MAE particles.

**2. Samples**

In this work, MAEs with ferromagnetic nickel microparticles with purity 0.999 were studied. The corresponding histogram of particle size distribution is shown in figure 1. The average size of particles was 15 μm. The image of Ni particles that were used when preparing the MAE is shown in the inset in figure 1.

The MAE matrix surrounding the particles was prepared from two-component high-strength and heat-resistant silicon with a density of 1.0-1.1 g/cm$^3$, a Shore A hardness of 22-26, and a relative elongation within an interval of 150-200%. The MAE density was $\rho = 2.45$ g/cm$^3$, and the volume concentration of nickel particles in the MAE was $f_1 = 0.15$ (i.e. 15 vol%).

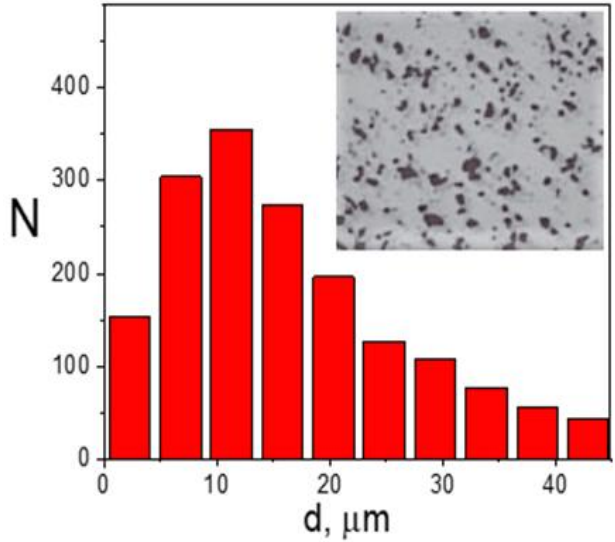

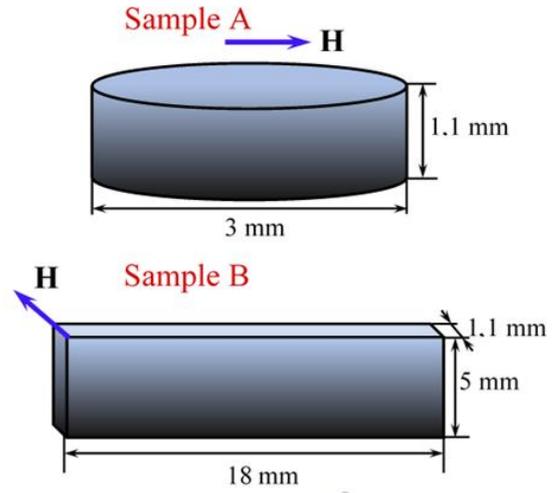

**Figure 1.** Histogram for the number *N* of nickel particles with the size *d*. The particle sizes were determined from the image of particles dispersed on a flat surface (see the inset).

The experimentally measured Young's modulus (*E*) of the researched MAE sample was found to equal 0.4 MPa. The value of *E* was determined from the dependence of strain on stress. According to the classification [53], the matrix of this MAE should be regarded as elastically stiff. The matrix with such elastic properties allowed us to study the bending of the console fabricated from the MAE and arranged horizontally rather than in a hanging position.

For magnetization measurements, disk-like sample A 3 mm in diameter and 1.1 mm in thickness was used (figure 2). The magnetization was measured on an LDJ-9500 magnetometer in a uniform magnetic field directed perpendicularly to the sample axis. The saturation magnetization at $T$ = 293 K was $m_S$ = 28.5 emu/g (or 70 emu/cm$^3$, or 70 kA/m). Such a reduction of the MAE saturation magnetization in comparison with that of bulk nickel (470 emu/cm$^3$) was proportional to the concentration of particles in the researched sample ($f_2$ = 70/470 ≈ 0.15).

The equality of two concentrations $f_1$ and $f_2$ confirms that the particles consisted of high purity nickel. The particles, as expected, had not changed their magnetic properties in the matrix, which confirms the absence of the matrix influence on the magnetic properties of particles.

To study the MAE bending in a uniform magnetic field and the shape memory effect, we used sample B fabricated from the MAE with the same composition as for sample A, but made in the form of a thin beam 18 mm in length and with a rectangular cross-section of 1.1×5 mm$^2$ (figure 2). When measuring the bending magnitude and studying the shape memory effect, one of the beam ends was fixed and the other remained free (the cantilever configuration, see photo in figure 2).

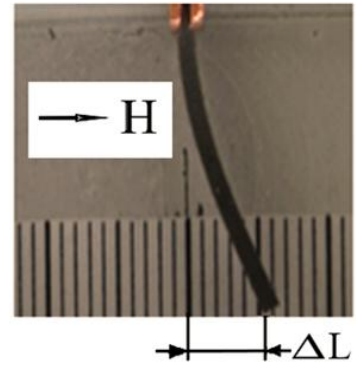

Figure 2. The magnetic field H is directed along the plane for sample A and perpendicular to the largest face for sample B. Photo: the top view of the beam bending Δ*L* induced by a perpendicular field *H*. The scale is graduated in millimeters.

Sample B was arranged horizontally between the poles of a dc electromagnet. Owing to a large diameter of electromagnet tips, the stationary magnetic field created by the electromagnet had a high degree of uniformity. It was directed horizontally and perpendicularly to the beam axis. When studying the bending, the magnetic field was directed perpendicularly to the largest face of the beam. Mechanical forces required to bend the beam in this direction are minimum. Additionally, the shape factor of the beam in this direction is the largest.

The direction perpendicular to the largest face of the beam is a direction of its hard magnetization. For this reason, the magnetization of the beam at its bending in a perpendicular field is not co-directional with the magnetic field. If the magnetization is not collinear with the magnetic field, then the beam is subjected to an action of a torque from the field side. It is this torque created by the magnetic field that makes the beam bend.If the magnetization is not collinear with the magnetic field, then the beam is subjected to a torque from the field. It is this torque generated by the magnetic field that creates the bending.

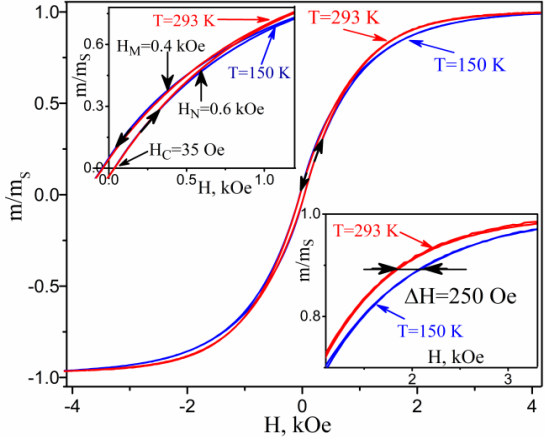

**Figure 3.** Field dependences of the normalized magnetization $m(H)/m_S(T)$ of MAE with nickel microparticles in silicon at $T$ = 293 K and 150 K. The sections of hysteresis loops at $H$ < 1 kOe and 1 kOe < $H$ < 3 kOe are shown in the upper-left and bottom-right insets, respectively.

## 3. Magnetic characteristics

Figure 3 demonstrates the field dependences of the normalized magnetization $m(H)/m_S(T)$ for MAE sample A both at room temperature $T$ = 293 K ($t = 20\,^\circ\text{C}$) and when cooled down to $T$ = 150 K ($t = -123\,^\circ\text{C}$). Here, $H$ is the magnetic field strength, $m(H)$ the projection of sample magnetization on the axis along which the magnetic field is directed, and $m_S(T)$ the saturation magnetization at the measurement temperature $T$. The magnetic field sweep rate equals 15 Oe/s. The downward arrow along the dependence curve indicates the decrease of the magnetic field, and the upward arrow its increase. The hysteresis loops differ little in their shape at $T$ = 293 K and $T$ = 150 K. The coercive force equals $H_c$ = 35 Oe (2.8 kA/m) (see the upper-left inset in figure 3). The residual magnetization is less than 4% of the saturation magnetization. Nickel microparticles are in a multi-domain ferromagnetic state.

From the bottom-right inset in figure 3, one can see that the magnetization has no hysteresis at the fields $1\,\text{kOe} < H < 3\,\text{kOe}$, and the magnetization curve corresponding to $T$ = 293 K lies a little higher than the magnetization curve corresponding to $T$ = 150 K. Before saturation, the shift of the curve obtained at $T$ = 293 K with respect to the curve obtained at $T$ = 150 K is about $\Delta H \approx 250$ Oe (see the bottom-right inset in figure 3).

The faster magnetization of MAE at $T$ = 293 K can be induced by a minor restructuring in the MAE, which favors its more rapid magnetization. As one can see from figure 3 (the bottom-right inset), the influence of MAE restructuring on the magnetization of sample A with an elastically stiff matrix is much weaker than in the MAE with elastically soft matrices [66]. However, the restructuring effect turned out significantly stronger at the magnetic-field-induced bending of sample B.

## 4. Bending of MAE by magnetic field

From the symmetry viewpoint, the inversion of the field in the equilibrium process should not be accompanied by a change in the direction (sign) of the bending value for a uniform MAE with ideally soft magnetic particles. In [51], it was theoretically obtained that such a symmetry property of magnetoelasticity can lead to a critically developing rotation of a magnetically anisotropic particle in the elastomer.

### 4.1 Field dependences for the bending value

In figure 4, the field dependences of the displacement of the free beam end are depicted. They were obtained in a sweep cycle when the magnetic field increased (the upward arrows) and decreased (the downward arrows). The sweep rate of the magnetic field was 10 Oe/s, the bending developed as slowly as the magnetization did (see figure 3). If the sweep rate was even slower, the $\Delta L(H)$ dependence remained almost unchanged. The saturation of the $\Delta L$ value was reached under a field 1.5 kOe.

The field dependence $\Delta L(H)$ has a hysteresis loop. In the backward sweep, when the magnetic field decreases, the corresponding $\Delta L(H)$ branch goes above the branch obtained for the increasing field.

When the magnetic field is turned off, the amount of residual bending $\Delta L(H = 0) = 0$. The loop for the field dependence of bending $\Delta L(H)$, in contrast to the loop for magnetization, has the "pinched" form.

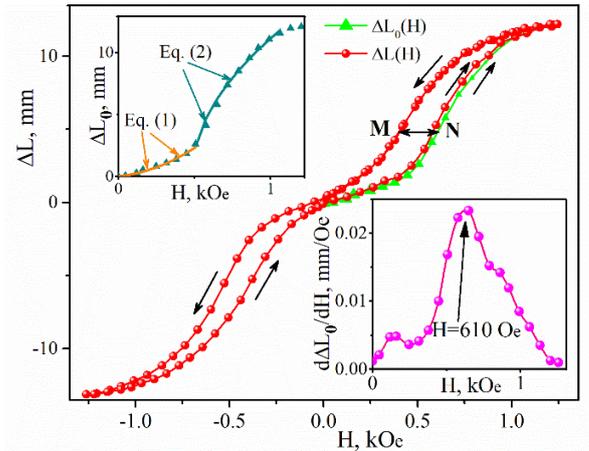

**Figure 4.** Field dependence of bending value $\Delta L(H)$. The initial magnetization curve is indicated as $\Delta L_0(H)$. Solid curves in the inset demonstrate the behavior of the dependence $\Delta L_0(H)$ obtained from equations (1) and (2). The field dependence for the derivative $d\Delta L_0(H)/dH$ is shown in the lower inset.

The initial curve $\Delta L_0(H)$ in figure 4 was obtained for an initially non-magnetized sample and the magnetic field growing from zero to the maximum positive value. For all other curves, owing to the availability of residual magnetization after the field decreasing to zero, the $\Delta L(H)$ depends on the magnetic field sign, and the bending value is strictly determined by the field direction. The beam axis is an easy magnetization axis. Therefore, when the magnetic field vanishes, the residual magnetization has a projection along the beam axis, and this projection of residual magnetization does not change its sign when a hysteresis loop in the $\Delta L(H)$ dependence is obtained.

An assumption can be made that the hysteresis in the dependence $\Delta L(H)$ originates from the hysteresis in the magnetization dependence $m(H)$. Owing to hysteresis, the magnetization is higher when the magnetic field decreases. Therefore, we may expect that the torque created by the magnetic field should be larger on the upper branch. As a result, owing to the hysteresis of $m(H)$, the curve $\Delta L(H)$ obtained for the decreasing magnetic field should lie above its counterpart obtained for the increasing magnetic field. Such a scenario is in qualitative agreement with the hysteresis exhibited in figure 3. However, the narrow hysteresis of $m(H)$ in the researched MAE was found to be not enough for the explanation of the wide hysteresis in the $\Delta L(H)$ dependence observed experimentally.

In figure 4, two points in the $\Delta L(H)$ dependence are selected as an example (their positions are indicated by a horizontal left-right arrow). These are point M (at $H_M = 0.4\,\text{kOe}$) on the upper hysteresis branch corresponding to the decreasing field and point N (at $H_N = 0.6\,\text{kOe}$) on the lower branch corresponding to the increasing field. The bending values are identical at those points, $\Delta L(H_N) = \Delta L(H_M)$. Therefore, the corresponding torques created by the magnetic field and acting on the beam must also be identical. To provide the equality between the torque magnitudes, the reduction of the field at point M has to be compensated by an increase of magnetization at this point.

However, from the upper-left inset in figure 3, one can see that the magnetization $m(H_M)$ at point M (in the field $H_M = 0.4\,\text{kOe}$) is 20% lower than its magnetization $m(H_N)$ at point N (in the field $H_N = 0.6\,\text{kOe}$). Instead of the expected inequality $m(H_M) \geq m(H_N)$, the experiment gives the opposite one, i.e. $m(H_M) < m(H_N)$. Hence, the hysteresis in the dependence $\Delta L(H)$ for the beam bending cannot be explained by the hysteresis in the magnetization dependence $m(H)$ shown in figure 3. Thus, the narrow hysteresis of the magnetization $m$ cannot be the main origin of the wide hysteresis in the bending $\Delta L$.

Hence, the specific features of the hysteresis in the dependence $\Delta L(H)$ are as follows: (i) the absence of residual deformation, which means that this hysteresis is not associated with relaxation processes; (ii) this hysteresis is not a consequence of the weak magnetization hysteresis, $m(H)$; and (iii) it is associated with the magneto-rheological effect.

The hysteresis in the dependence $\Delta L(H)$ is induced by a modification in the elastic properties of the MAE due to the restructuring of the latter in the course of its bending under the action of the magnetic field. If the bending is large, the filler particles become displaced with respect to one another. When strongly magnetized particles approach, they can touch one another owing to the magnetic force action between them, as was described in [67,68]. Two magnetized particles mutually enhance the magnetization of each other, so their interaction becomes strongly nonlinear at short distances. As a result of the nonlinearity of interactions between the particles, a state with touching magnetized particles turns out more stable when the external magnetic field decreases [68]. A similar situation with the hysteresis of bending in a system of magnetic stripes on an elastomer was considered in [43]. In that case, the touching magnetized stripes attempted to sustain bending, which led to hysteresis. Thus, the hysteresis in the bending of sample B induced by the action of an external magnetic field is associated with the restructuring in the MAE at its bending and is a consequence of the magneto-rheological effect.

### 4.2. Critical behavior of the bending

Before obtaining the initial curve, which is shown in the upper inset in figure 4, the sample was not exposed to a magnetic field. This curve $\Delta L_0(H)$ is the best fit for the equilibrium process. The lower branch of the hysteresis loop $\Delta L(H)$ runs slightly above the curve. The dependence is continuous and has two sections: a low-field one with a slowly increasing bending value and a high-field one with a fast increase in the bending value and its value reaching saturation.

The inset in figure 4 demonstrates that the behavior of $\Delta L_0(H)$ in the interval $0 < H < 0.5\,\text{kOe}$ can be approximated by the dependence

$$\Delta L_0(H) = a_1 H + a_2 H^2 \qquad (1)$$

with $a_1 = 1.3\,\text{mm/kOe}$ and $a_2 = 6.5\,\text{mm/kOe}^2$. In the field interval $0.5\,\text{kOe} < H < 1\,\text{kOe}$, i.e. up to the bending saturation, a good approximation for $\Delta L_0(H)$ (see the inset in figure 4) looks like

$$\Delta L_0(H) = b(H - H_{cr})^\gamma, \qquad (2)$$

where $b = 15.6\,\text{mm}/\text{kOe}^{0.51}$, $\gamma = 0.51$, and $H_{cr} = 0.48\,\text{kOe}$. Hence, from (3), it follows that at $0.5\,\text{kOe} < H < 1\,\text{kOe}$, the field dependence of the bending magnitude is described by a power-law dependence with the power exponent $\gamma \approx 0.5$. In a similar way and with the same power exponent $\alpha = 0.5$ evolves the critical rotation of a magnetically anisotropic ferromagnetic particle induced by a magnetic field in an elastomer [51].

The inset in figure 4 shows the field dependence for the derivative $d\Delta L_0/dH$, which has a sharp maximum at a magnetic field of about 0.61 kOe. Such a rapid increase is absent in the field dependence of magnetization (see figure 3).

In the low-field region $H < 0.45$ kOe, the value of the derivative $d\Delta L_0/dH$ is small and its changes are insignificant. At $H > 0.45$ kOe, it begins to increase sharply to a maximum, after which its rapid decline begins, and only then the saturation of the value $\Delta L_0(H)$ is observed, when the derivative $d\Delta L_0/dH$ again becomes small. The region of the derivative maximum turns out to be narrow, with the half-width at half-maximumamplitude being equal to 0.2 kOe.

Thus, the critical behavior for a continuously changing quantity $\Delta L_0(H)$ is characterized by a singularity in the field dependence of its derivative $d\Delta L_0/dH$. The behavior of this feature is similar to the behavior of the magnetization and its derivatives during the phase transition of the second order induced by a magnetic field [69]. In the theory of phase transitions, the derivatives of the observed physical quantities at the critical point can tend to infinity, but in experiments they are usually always finite due to structural inhomogeneities.

A critical bending occurs when the magnetostatic and elastic energies of a sample compete. To estimate the critical field $H_{cr}$ theoretically, let us analyze the magnetostatic and elastic energies in the same way as was done in [43,54,55]. Let us assume that the magnetic field is uniform and the magnetic properties of the MAE are also uniform. As a result, we obtain an expression for $H_{cr}$, which can be written as follows (see Appendix A):

$$H_{cr} = \frac{\sqrt{E}}{\chi}\frac{\delta}{l}\sqrt{\frac{1+4\pi\chi}{16\pi}}, \qquad (3)$$

where $\chi$ is the magnetic susceptibility of the MAE, $l$ is the length of the largest side of the beam, $\delta$ is the length of its smallest side. Substituting the sample parameters into (3), we obtain $H_{cr} \approx 0.34$ kOe. This theoretical estimation for the critical field $H_{cr}$ is in good qualitative agreement with the experimental data.

The theory gives that if $H < H_{cr}$, there is no bending ($\Delta L = 0$), and if $H > H_{cr}$, the bending value is proportional to the square root, $\Delta L \sim \sqrt{H - H_{cr}}$ (see Appendix A). This dependence brings about the critical exponent $\gamma = 1/2$, which is in good agreement with (2). For such a field dependence $\Delta L(H)$, its derivative has a singularity with an infinitely steep increase at $H \to H_{cr}$ when $H < H_{cr}$, and a slower decay of its value at $H > H_{cr}$. In the experiment, the accelerated increase in the derivative is shown in the inset in figure 4.

The critical behavior of the bending is associated with a state symmetry change. If $H < H_{cr}$, the beam is magnetized uniformly on average and the beam magnetization vector is codirectional with the magnetic field strength vector, $\mathbf{m} \uparrow\uparrow \mathbf{H}$. But if $H < H_{cr}$, the magnetization is directed perpendicularly to the beam. This is a high-symmetric magnetic state free of torque, and the beam is not deformed. In the case when $H > H_{cr}$ and $(H-H_{cr})/H_{cr} << 1$, the magnetization vector $\mathbf{m}$ becomes noncollinear to the magnetic field strength vector $\mathbf{H}$ at every point in the beam. This is a low-symmetric magnetic state in which the beam is subjected to the action of a torque created by the magnetic field. Hence, the beam bending and the noncollinearity of the vectors $\mathbf{m}$ and $\mathbf{H}$ comprise a mutually consistent and nonlinear process.

Thus, at the critical point $H = H_{cr}$, a magnetic phase transition takes place from a high-symmetric magnetic state into a low-symmetric one. The critical bending is a consequence of the change in the symmetry of the beam magnetic state, which is similar to those described in the Landau theory of phase transitions.

Thus, a uniform magnetic field induces a critical bending of the MAE, if the matrix is highly elastic, the filler particles are easily magnetized, and the MAE, as a composite, has uniform magnetic properties.

Note that the viscosity of the MAE elastomer can affect the critical bending. However, it should be taken into account that viscosity and dissipation only smooth out the course of any critical process. If the viscosity is high, there is no critical bending at all. Thus, our choice of an elastically stiff matrix made it possible to avoid the negative effect of viscosity, and its action did not blur the critical bending behavior (see figure 4).

## 5. Shape memory effect

From the data presented in figure 5, it follows that the magnetic field can induce a significant bending in an MAE with an elastic-rigid matrix. This fact testifies that the thermo-magnetic shape memory effect [42] accompanied by deformation features at critical temperatures of the matrix [60] is possible in the researched MAE sample. In MAEs, unlike polymers (see figure 4), bending deformation can be created by a magnetic field without applying external mechanical stress.

As a rule, the shape memory effect in polymers is illustrated with the help of plots describing relationships between the magnitudes of mechanical stresses, strains and the temperature [70-72]. For MAEs, the magnetic forces play the role of a deformation source. Therefore, the magnetic field strength can be one of the parameters for the plots illustrating the magnetic shape memory in MAEs. A schematic diagram beam bending versus temperature for sample B is shown in figure 5. At first, the stationary uniform magnetic field $H$=0.6 kOe was applied perpendicularly to the initially undeformed sample at room temperature. Under the action of this field, the sample transited into a bent state (path $\text{I} \to \text{II}$ in figure 5). Then, the sample was cooled down in liquid nitrogen keeping the magnetic field constant (path $\text{II} \to \text{III}$ in figure 5). In state III, the magnetic field was switched off, $H = 0$, but the sample remained to be bent. Finally, the sample was heated up to room temperature in the absence of the magnetic field (path $\text{III} \to \text{I}$ in figure 5).

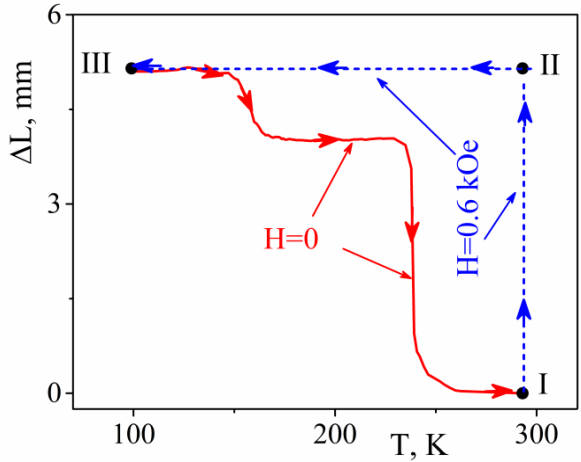

**Figure 5.** Beam bending, $\Delta L$, versus temperature, $T$, diagram measured for sample B. Point I corresponds to the initial state of the sample at room temperature; path $\text{I} \to \text{II}$ to the sample bending in the field $H = 0.6\,\text{kOe}$ at room temperature; path $\text{II} \to \text{III}$ to the sample cooling in the field $H = 0.6\,\text{kOe}$; at point III, the magnetic field is switched off, $H = 0$; path $\text{III} \to \text{I}$ corresponds to the sample heating to room temperature at $H = 0$.

The closed loop $\text{I} \to \text{II} \to \text{III} \to \text{I}$ in the diagram is nothing else but the shape memory effect [72,73]. The MAE undergoing the magnetic-field-induced bending, when being cooled down, retained its bending even after the magnetic field had been completely switched off. However, after being heated up, the MAE "restored" its initial shape.

In path $\text{III} \to \text{I}$ of the loop (here, the sample is heated up to room temperature at $H = 0$), there are two intervals where the rate of the beam bending reduction is substantially higher. Within the interval $148\,\text{K} < T < 165\,\text{K}$, a rapid but continuous decrease of the bending magnitude by almost 20% takes place. This temperature interval corresponds to the matrix glass transition [66,74].

At $T$=238 K, the beam demonstrates an abrupt change of its bending with an almost complete return to the shape of the initially undeformed sample at room temperature. This jump in the dependence $\Delta L(T, H = 0)$ is associated with the matrix melting/solidification at $T$=238 K, a jump-like process accompanied by the change of particle mobility in the MAE [52,66,74]. At a temperature lower than the solidification one, the positions of the magnetic particles in the matrix become fixed (effect of blocking), so that the magneto-rheological effect of MAE diminishes [52,66]. In [52,66], when studying the effect of blocking, the sample was placed in an elastic-rigid cuvette and, therefore, the sample did not deform during magnetization. Figure 4 shows the bending measurement data for a free sample, when its size and shape were not limited by the cuvette, as was done in [52,66].

When the temperature drops, the elastic moduli of the polymer matrix increase. However, the growth of the elastic moduli does not make it possible to explain the shape memory effect and the diagram exhibited in figure 5. If there were no melting/solidification and glass transition, then, after the magnetic field is turned off at point III, the sample should straighten out and the bending value should become zero, $\Delta L(T = 100K, H = 0) = 0$. An increase in the value of the elastic modulus will additionally contribute to the straightening of the sample after the magnetic field is switched off. In the experiment, after the magnetic field had been switched off, there was no straightening at point III and the bending survived when the magnetic field hadbeen switched off.

Similarly to the case of polymers [60], the shape memory effect in MAEs is associated with the melting/solidification or the glass transition of the matrix rather than the change or relaxation of their elastic properties. Obviously, in MAEs, the matrix does not affect the magnetic properties of the particles, and the particles do not have a significant effect on the melting/solidification or glass transition of the matrix. It was shown in [66] experimentally using the differential scanning calorimetry method. The diagram in figure 6 is another direct proof of that statement.

## 6. Threshold field for the magnetic shape memory effect in the MAE

The MAE can be heated above its glass transition and melting temperatures with the help of an ac magnetic field [21,25,27,61,62]. It is clear that heating can be carried out, for example, by blowing hot air over the MAE. However, in the case of soft robotics or medical applications, a direct contact with a heat source can be technologically limited or completely excluded. For such applications, heating must be performed contactlessly in an ac magnetic field, as it was suggested for the magnetic shape memory effect in MAEs (see [42]).

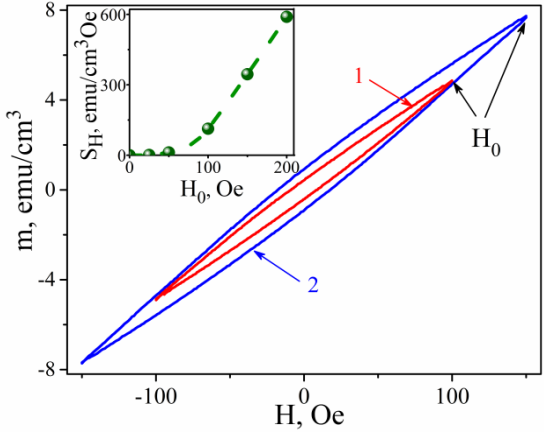

**Figure 6.** Partial hysteresis loops in the cases $H_0 = 100\,\text{Oe}$ (curve 1) and $H_0 = 150\,\text{Oe}$ (curve 2). The dependence of the loop area on the maximum field $H_0$ reached in the partial loop is shown in the inset.

As follows from [64], the MAE heating depends on the field amplitude, and this dependence has a threshold character. Namely, the MAE heating takes placed in an ac magnetic field only if the field amplitude is larger than a certain value. The origin of this threshold has not been clarified yet. It was shown in [65] that heating with the help of magnetic losses in an ac field can be estimated by analyzing the partial hysteresis loops. To study the mechanism of the amplitude threshold appearance for the MAE heating, we measured the partial loops at various sweep amplitudes of the magnetic field.

For example, in section $\text{III} \to \text{I}$ of the diagram exhibited in figure 5 (here, the stationary magnetic field is turned off, $H = 0$), the sample can be heated owing to the action of a small-amplitude ac magnetic field, similarly to what was done for magnetic suspensions [65]:

$$h = h_0 \cos 2\pi\nu t , \qquad (4)$$

where $h_0$ is the field amplitude, and $\nu$ is the frequency. The heating of MAE in a small-amplitude ac field can also have a threshold for the field amplitude, similarly to what happens at the heating of magnetic suspensions [75].

Figure 6 shows partial magnetization loops for sample A in the cases $-100\,\text{Oe} < H < 100\,\text{Oe}$ (loop 1) and $-150\,\text{Oe} < H < 150\,\text{Oe}$ (loop 2). The dependence of the loop area on the maximum reached field for the partial sample magnetization is shown in the inset.

The specific power losses per unit mass of MAE, $W$, in a low-frequency ac magnetic field can be evaluated from the expression

$$W = \frac{\nu S_H}{\rho} , \qquad (5)$$

where $\nu$ is the ac field frequency, $\rho$ the MAE density, and $S_H$ the hysteresis loop area (it is equal to the energy density absorbed by the MAE particles),

$$S_H = \oint_{(H)} m(H) dH . \qquad (6)$$

If the ac field frequency is $\nu = 300\,\text{kHz}$, the specific powerlosses equal $W = 1.38\,\text{kW/kg}$ for the field amplitude $h_0 = 100\,\text{Oe}$, being almost three times as much, $W = 5.5\,\text{kW/kg}$, for the field amplitude $h_0 = 150\,\text{Oe}$. Those values are slightly lower than the values obtained for MAEs with superparamagnetic particles [62].

From the inset in figure 6, one can see that the magnitude of magnetic losses for the examined weakly coercive MAE with nickel microparticles depends in a threshold manner on the maximum field amplitude $H_m$. Magnetic losses become substantial if the amplitude of ac field more than twice exceeds the coercive force.

Thus, the shape memory effect of the MAE in the small-amplitude ac magnetic field has a threshold character. The threshold value is associated with specific power losses and depends on the coercive force parameter of the ferromagnetic particles.

## 7. Conclusions

It was found that the MAE consisting of weakly coercive ferromagnetic particles embedded in an elastically rigid matrix exhibits a magneto-rheological effect with a strong hysteresis for the beam bending magnitude. The corresponding hysteresis loop has a "pinched" shape. It was shown that the bending magnitude critically depends on the field magnitude. The critical bending behavior of the MAE is symmetrical and characterized by the maximum rate of bending variation at the critical point. The critical bending is observed if the MAE as a composite has uniform magnetic properties, a high magnetic susceptibility, and is located in a uniform magnetic field. The critical bending is a property of the MAE material in whole rather than its components and can be observed in a lot of other MAEs with other compositions that meets the indicated conditions.

The critical bending arises as a magnetoelastic phenomenon accompanied by a change in the magnetic state symmetry, as it occurs in the Landau theory of phase transitions. In the high-symmetric state, when the magnetization is directed along the magnetic field, there is no torque and the beam is not bent. In the low-symmetric state, when the vector of beam magnetization and the vector of the magnetic field strength are oppositely directed, there arises a torque, which bends the beam. The beam bending and magnetization comprise a self-consistent and non-linear process.

A weakly coercive MAE can be heated up by applying a low-amplitude ac magnetic field, which invokes magnetic losses. The magnitude of losses depends non-linearly on the field amplitude. For heating to occur, the amplitude of the ac field should exceed the coercive force $H_C$. When the amplitude of the magnetic field is less than the coercive force, then it weakly magnetizes the MAE particles and, accordingly, the loop area is also small. When the magnetic field is greater than the coercive force, the magnetization of the particles increases, and the loop area and the specific power losses increase too. Therefore, the MAE heating depends nonlinearly on the amplitude of the ac magnetic field.

The shape memory effect was revealed for the bending of a MAE beam in a cycle consisting of the magnetization, cooling in the magnetic field, and heating in the absence of magnetic field. It was shown that the magnetic shape memory effect in MAEs may depend on the glass transition and melting/solidification temperatures of the matrix. In this case, the deformation of the beam and its heating can be performed by subjecting the sample to the contactless action of the external dc and ac magnetic fields, which is important when creating shape memory devices for soft robotics or medical applications. Note that the temperature at which the shape memory effect occurs can be changed by selecting the elastomer matrix and magnetic particles. For example, to obtain the magnetic shape memory in the MAE at room temperature, an elastomer with the glass transition or melting/solidification temperature higher than room temperature but lower than the Curie temperature of the filler magnetic particles has to be used.

### Acknowledgements

This work was partially supported by "The Thousand Talents Program for Foreign Experts" (Project WQ20162200339).

### Appendix A. Theory of critical bending

Let us consider a thin beam, $\delta/l \ll 1$, with a small bending, i.e. when $l/R \ll 1$, where $\delta$, $l$, and $R$ are the thickness and the length of the beam and the curvature radius of the beam bending, respectively (see figure A1). The curve D in figure A1 denotes a curve the length $l$ of which does not change at bending, i.e. $l$=const. Along other curves, the sample becomes stretched or compressed. Let us introduce a local coordinate $\xi$ to describe the point position in the MAE. Its value is equal to the smallest distance from this point to the curve D. Now, let us consider an elementary section $dl(\xi=0)$ of the curve D. The section located at the distance $\xi$ from the curve D has the length $dl(\xi)$. There is a simple relationship between the lengths of those sections: $dl(\xi)/(R+\xi) = dl(\xi=0)/R$. The relative deformation equals $\varepsilon(\xi) = (dl(\xi) - dl(\xi=0))/dl(\xi=0) = \xi/R$. The elasenergy is determined using the integration procedure

$$U_e = \int_V \frac{1}{2} E\varepsilon^2 dV = V \frac{E}{24} \frac{\delta^2}{l^2} \alpha^2, \quad (A1)$$

where $V$ is the beam volume, and $\alpha = l/R$ is the angle characterizing the bending curvature (see figure A1): if $\alpha=0$, then $R \to \infty$ and the bending is absent; otherwise, if $\alpha \neq 0$, then $R$ has a finite value, i.e. the beam is bent.

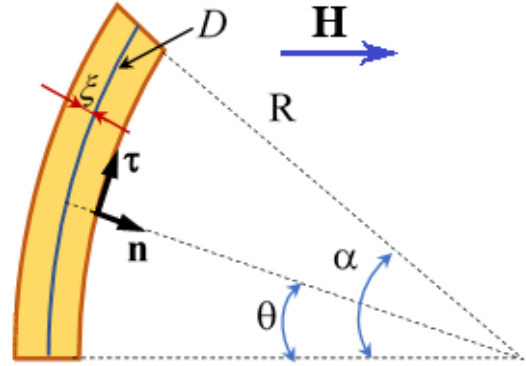

**Figure A1**. Beam bending in the magnetic field **H**.

Let us write the magnetic energy in the form

$$U_m = -\sum_{j=1}^{N} \int_{V_j} (\mathbf{m}_j \mathbf{H} + \frac{1}{2}\mathbf{m}_j \mathbf{h}_j^m) dV_j, \quad (A2)$$

where $\mathbf{m}_j$ and $\mathbf{h}_j^m$ are the magnetization vector and the magnetostatic field, respectively, for the $j$-th magnetic particle in the MAE matrix; **H** is the external uniform magnetic field, and integration is carried out over the volume $V_j$ of the $j$-th magnetic particle. Hereafter, we suppose the particles to have a shape close to ellipsoidal. We also suppose that the particles are located at large distances from one another, which allows their internal magnetic fields to be considered as uniform. Then, there is an evident relationship between the magnetization vector and the field components in the $j$-th particle,

$$\mathbf{m}_j = \chi_0 (\mathbf{H} + \mathbf{h}_j^m), \quad (A3)$$

where $\chi_0$ is the magnetic susceptibility of the substance that the ferromagnetic particle is made of.

For a magnetically soft material, $\chi_0 \gg 1$. Therefore, we have $\mathbf{h}_j^m \approx -\mathbf{H}$ inside the particle. Then, expression (A2) can be simplified to the form

$$U_m = -\frac{1}{2}\sum_{j=1}^{N} V_j \mathbf{m}_j \mathbf{H}. \qquad (A4)$$

In the continuum approximation, formula (A4) can be rewritten as the integral

$$U_m = -\frac{1}{2}\int_V \mathbf{m}\mathbf{H}dV, \qquad (A5)$$

where $\mathbf{m}(\mathbf{r}) = \frac{1}{\Delta V}\sum_{j=1}^{\Delta N} V_j \mathbf{m}_j$ is the average MAE magnetization at the point **r**, and $\Delta N$ is the number of particle in the volume $\Delta V$ (we assume that $\Delta N \gg 1$ and $\Delta V \ll \delta^3$).

For a thin beam, the demagnetizing field equals $\mathbf{h}^m = -4\pi N_{\delta\delta}(\mathbf{mn})\mathbf{n}$, where $N_{\delta\delta} \approx 1$ is the shape factor along the direction transverse to the beam, and **n** is a unit vector perpendicular to the beam.

Thus, in the case of low fields, when the relationship between the beam magnetization and the field is linear, we may write

$$m_\parallel = \chi \mathbf{H}\boldsymbol{\tau} = \chi H \sin\theta, \; m_\perp = \frac{\chi \mathbf{Hn}}{1+4\pi\chi} = \frac{\chi H \cos\theta}{1+4\pi\chi}, \quad (A6)$$

where $\chi$ is the magnetic susceptibility of the MAE, $\boldsymbol{\tau}$ is a unit vectors directed along the beam, $m_\parallel$ and $m_\perp$ are the projections of the magnetization vector ($\mathbf{m} = m_\parallel \boldsymbol{\tau} + m_\perp \mathbf{n}$), and $\theta$ is the angle between the vectors **H** and **n** (see figure A1).

Integrating (A5) using (A6), we obtain the following expression for the magnetic energy:

$$U_m = -Vu_m\left(\frac{1}{2\pi\chi} + \left(1 - \frac{\sin 2\alpha}{2\alpha}\right)\right), \qquad (A7)$$

where $u_m = \pi(\chi H)^2/(1+4\pi\chi)$ is a field-dependent parameter.

Assuming that the bending is small, let us express the total energy $U = U_e + U_m$ as a series expansion in the small parameter $\alpha$ and confine it to the forth-order term,

$$U = -\frac{u_m}{2\pi\chi}V - \frac{2}{3}u_m V\left(\left(1 - \frac{E}{16u_m}\frac{\delta^2}{l^2}\right)\alpha^2 - \frac{1}{5}\alpha^4\right). \quad (A8)$$

This expression contains only even power exponents of $\alpha$, which is a requirement of symmetry.

By minimizing (A8) with respect to $\alpha$, from the condition $\partial U/\partial \alpha = 0$, we obtain two possible solutions,

$$\alpha = 0 \; \text{and} \; \alpha = \pm\sqrt{\frac{5}{2}\left(1 - \frac{E}{16u_m}\frac{\delta^2}{l^2}\right)}. \qquad (A9)$$

The solution $\alpha = 0$ exists for all magnetic field values. It is stable if $u_m < E\delta^2/(4l)^2$. The other solution is possible and stable at $u_m > E\delta^2/(4l)^2$. From (A9) one can see that the bending develops critically as a bifurcation in the critical field

$$H_{cr} = \frac{\sqrt{E}}{\chi}\frac{\delta}{l}\sqrt{\frac{1+4\pi\chi}{16\pi}}. \qquad (A10)$$

At small bending deformations, the experimentally measured quantity $\Delta L$ can be written in the form $\Delta L = l\alpha/2$. Using (A8), it is easy to show that $\Delta L$ also critically depends on the magnetic field. In particular, in the closest vicinity of $H_{cr}$, the field dependence $\Delta L(H)$ looks like

$$\Delta L(H) = \begin{cases} 0, & H < H_{cr} \\ l\sqrt{\frac{5}{4}\frac{H-H_{cr}}{H_{cr}}}, & H > H_{cr} \end{cases}. \qquad (A11)$$

If $H < H_{cr}$, then the undeformed state ($\Delta L(H) = 0$) is stable, in which the magnetization vector is collinear to the field, $\mathbf{m} \uparrow\uparrow \mathbf{H}$; this is a high-symmetric magnetic state. In this state, the torque equals zero and the beam does not undergo bending. If $H > H_{cr}$, then the beam state with the bending $\Delta L(H) \sim l\sqrt{H-H_{cr}}$ is stable, in which the vectors **m** and **H** are noncollinear; this is a low-symmetric magnetic state. In the low-symmetric magnetic state, the torque does not equal zero, so the beam is bent. The magnetization and the bending of the beam comprise a mutually consistent process, which develops critically, similarly to those described in the Landau theory of phase transitions.

**ORCID iDs**

V M Kalita  http://orcid.org/0000-0001-6329-9095

G G Levchenko  http://orcid.org/0000-0003-2287-8626